\documentclass[11pt]{article}
\usepackage{euscript}
\usepackage{amssymb}
\usepackage{amsfonts}
\usepackage{amsbsy}
\usepackage{epsfig}
\usepackage{amsthm}
\usepackage{amscd}
\usepackage{amstext}
\usepackage{verbatim}
\usepackage{amsmath}
\usepackage{hyperref}
\usepackage{cite}

\pdfoutput=1

\begin{document}

\begin{center}
{\Large \textbf{Holographic complexity of the disk subregion in
(2+1)-dimensional gapped systems }}

\vspace{1cm}

Lin-Peng Du$^{1}$, Shao-Feng Wu$^{1,2,3}$, Hua-Bi Zeng$^{2,4}$

\vspace{1cm}

{\small \textit{$^{1}$Department of Physics, Shanghai University, Shanghai,
200444, China }}\\[0pt]
{\small \textit{$^{2}$Center for Gravitation and Cosmology, Yangzhou
University, Yangzhou 225009, China }}\\[0pt]
{\small \textit{$^{3}$The Shanghai Key Lab of Astrophysics, Shanghai,
200234, China}}\\[0pt]
{\small \textit{$^{4}$College of Physics Science and Technology, Yangzhou
University, Yangzhou 225009, China}}\\[0pt]
\vspace{0.5cm} {\small \textit{dulp15@shu.edu.cn, sfwu@shu.edu.cn,
zenghbi@gmail.com}}
\end{center}

\vspace{1cm}

\begin{abstract}
Using the volume of the space enclosed by the Ryu-Takayanagi (RT) surface,
we study the complexity of the disk-shape subregion (with radius R) in
various (2+1)-dimensional gapped systems with gravity dual. These systems
include a class of toy models with singular IR and the bottom-up models for
quantum chromodynamics and fractional quantum Hall effects. Two main results
are: i) in the large-R expansion of the complexity, the R-linear term is
always absent, similar to the absence of topological entanglement entropy;
ii) when the entanglement entropy exhibits the classic `swallowtail' phase
transition, the complexity is sensitive but reacts differently.
\end{abstract}
\pagebreak

\section{Introduction}

Holographic principle suggests that the spacetime could be an emergent
phenomenon. Entanglement, which characterizes how the joint of different
parts of a system distinguishes its classical whole, is believed to play an
essential role \cite{Swingle0905,Raamsdonk0907}. This insight is due in
large part to the holographic prescription for the entanglement
entropy---the area of the codimension-two minimal surface in the bulk with
its UV boundary coincident with the entangling surface in the dual field
theory \cite{RT}. The mentioned minimal surface has been referred as the
Ryu-Takayanagi (RT) surface.

Complexity is another quantum information quantity that might be important
in understanding the quantum structure of spacetimes \cite{Susskind1411}. It
is relevant to the number of unitary operators which converts one quantum
state to another. There are two holographic dualities which have been
proposed to describe the complexity. Popularly, they are known as the
`complexity=volume' (CV) conjecture \cite{Susskind1403} and
`complexity=action' (CA) conjecture \cite{Brown1509}. Starting from the
study of the linear growth of the black hole formation and the size of the
Einstein-Rosen bridge, the former conjecture states that the complexity in
the boundary field theory is related to the volume of a codimension-one
maximal bulk space, which is anchored on a given boundary time slice. The CV
conjecture needs to introduce a length scale which depends on the concrete
systems. Without such arbitrariness, the latter conjecture states that the
complexity is described by the bulk action on the Wheeler-DeWitt patch \cite%
{Brown1509}, which is the domain of dependence of any Cauchy surfaces in the
bulk that approaches the boundary time slice.

In this paper, we will be concerned with the RT volume that is associated
with the bulk space enclosed by the RT surface. Intuitively, the RT volume
is interesting since it is intrinsically related to the holographic
entanglement entropy (HEE). The lesson learned from the RT volume would
suggest new holographic duals to the quantum information. In fact, motivated
by the connection between the volume of the maximal time slice in an Anti-de
Sitter (AdS) spacetime and the fidelity susceptibility of pure states \cite%
{Miyaji1507}, the RT volume was proposed as the holographic subregion
complexity (HSC), corresponding to the reduced fidelity susceptibility of
mixed states in the boundary \cite{Alishahiha1509,Carmi1609}. Furthermore,
the quantitative evidence of this correspondence has been presented by
studying the marginal perturbation in the (1+1)-dimensional conformal field
theory (CFT) \cite{Shu1702}. Moreover, for a spherical subregion in the
boundary, it was also argued that the regularized RT volume is related to
the Fisher information metric \cite{Banerjee1701}. On the other hand, the
HSC has a deep relationship to the HEE indeed. For example, similar to the
HEE, the HSC can signal different phase transitions \cite%
{Mazhari1601,Roy1701,Wang1704}. Besides the aforementioned work, the RT
volume has attracted much attention, e.g., see \cite%
{Myers1612,Momeni1604,Bakhshaei1703,Karara1711,Chen1803}. Among others, we
only note that the scale arbitrariness in the CV conjecture could be
eliminated by introducing some variant of the RT volume: it was proposed
recently within the AdS$_{3}$/CFT$_{2}$ duality that the curvature integral
in the space enclosed by the RT surface provides a dimensionless quantity
which could be dual to the complexity of the reduced density matrix \cite%
{Rene1710}.

For condensed matter theories (CMT), the quantum entanglement provides
important insight into the structure of many-body states. One famous example
is the notion of topological entanglement entropy (TEE) \cite%
{Kitaev0510,Wen2006,Grover1108}, which is one of the most proximate
representations of the topological order in (2+1)-dimensional topological
field theories. The TEE can be extracted by the asymptotic behavior of the
entanglement entropy between a disk (with radius R) and the rest of the
system: $\mathcal{S}=\alpha R-\mathcal{S}_{\mathrm{top}}+\mathcal{O}(R^{-1})$%
. Here the first term denotes the well-known area law and the second term is
the TEE. When the theory has a mass gap, the TEE is invariant under a
continuous deformation of the disk.

In \cite{Pakman0805}, the TEE is studied based on the RT prescription. It is
found that the TEE is vanishing for the AdS soliton, as expected for a
theory like quantum chromodynamics (QCD). The vanishing of the holographic
TEE is not accidental, which also can be seen in a class of
(2+1)-dimensional gapped geometries with singular IR \cite{Liu1202,Liu1309}.
The nonvanishing TEE can be produced by introducing the Chern-Simons
interaction in the D3-D7 systems \cite{Takayanagi0901} or the Gauss-Bonnet
(GB) curvature in the gravity theory when the RT surface has the disk
topology \cite{Parnachev1504}. Moreover, the TEE can be interpreted as the
black hole entropy in the AdS$_{3}$ \cite{Verlinde1308,Luo1709}.

A natural question is whether there is a quantity in the HSC, which
corresponds to the TEE in the HEE. As a first step to address this question,
we will study the large-R expansion of the HSC (with a disk-shape subregion)
in various (2+1)-dimensional holographic gapped systems without the TEE. On
the other hand, in some of these systems, the HEE can exhibit a classic
`swallowtail' phase transition when the topology of the RT surface changes
from the disk shape to the cylinder shape. This typical phenomenon has been
taken as a probe of the confinement/deconfinement transition \cite%
{Klebanov0709,Tatsuma0611,Pakman0805}. We will study the HSC in the
intermediate-R region and focus on whether the HSC is sensitive to the
topology change of the RT surface.

The rest of this paper is organized as follows. In Sec. 2, we will study a
class of gapped geometries%
\begin{equation}
ds^{2}=\frac{L^{2}}{z^{2}}\left[ -h(z)dt^{2}+d\rho ^{2}+\rho ^{2}d\theta
^{2}+\frac{dz^{2}}{f(z)}\right] ,  \label{metriczn}
\end{equation}%
where $L$ is the AdS radius and the IR behavior $\left( z\rightarrow \infty
\right) $ is required to be%
\begin{equation}
f(z)=az^{n}+\cdots ,\quad a>0,\;n\geq 2.  \label{fzn}
\end{equation}%
Then the spacetime becomes singular in the IR. We will prove analytically
that in the large-R expansion, the R-linear term is absent in the HSC.
Moreover, we will calculate numerically the HSC as a function of R in
several toy-models. It will be shown that the HSC can probe the topology
change of the RT surface. From Sec. 3 to Sec. 5, we will study three kinds
of (2+1)-dimensional phenomenon models with the mass gap, including the AdS
soliton, soft-wall model and holographic fractional quantum Hall (FQH)
model. For all the models, the R-linear term in the HSC is vanishing. For
the AdS soliton, the HSC is sensitive to the topology change. In Sec. 6, we
will discuss whether the R-linear term can be produced in the soft-wall
model by involving the GB correction to the gravity theory. In Sec. 7, the
conclusion will be given. In Appendix A, we will review briefly the
numerical solution of the holographic FQH model. In Appendix B, we attempt
to explore further the HSC in the GB gravity.

\section{Geometries with singular IR}

Suppose that a spacetime is described by the metric (\ref{metriczn}) with
the IR behavior (\ref{fzn}). When $n>0$, the singularity lies at a finite
proper distance away. From the UV/IR connection, it might be excepted that
the corresponding IR phase is gapped. Actually, by analyzing the spectrum of
a probed scalar field in the spacetime, it has been found \cite%
{Liu1202,Liu1309} that the system with $n>2$ has a discrete spectrum and for
$n=2$, it has a continuous spectrum above the gap. For $0<n<2$, the geometry
is singular but is not dual to a gapped phase.

\subsection{RT surface}

We will study the RT surfaces in those gapped geometries. Consider a circle
with radius R in the UV boundary, which divides the system into two parts.
The HEE between them can be determined by%
\begin{equation}
\mathcal{S}=\frac{\mathrm{Area}\left( \Sigma \right) }{4G_{N}},  \label{S0}
\end{equation}%
where $G_{N}$ is the four-dimensional Newton constant and $\Sigma $ denotes
the minimal surface that extends from the boundary circle to the bulk. Using
Eq. (\ref{metriczn}), one can read the induced metric of the RT surface:%
\begin{equation}
ds_{\mathrm{ind}}^{2}=\frac{L^{2}}{z^{2}}\left[ (\frac{1}{f}+\rho ^{\prime
2})dz^{2}+\rho ^{2}d\theta ^{2}\right] ,  \label{induce1}
\end{equation}%
where the prime denotes the derivative with respect to $z$ that is taken as
one of surface coordinates. By the induced metric, one can obtain the
entropy functional of $\rho (z)$:
\begin{equation}
\mathcal{S}=\frac{\pi L^{2}}{2G_{N}}\int_{0}^{\infty }dz\frac{\rho }{z^{2}}%
\sqrt{\frac{1}{f}+\rho ^{\prime 2}}.  \label{Area1}
\end{equation}%
Taking the variation with respect to $\rho (z)$, one can derive the equation
of motion (EOM) that determines the extremal surface:%
\begin{equation}
\rho ^{\prime \prime }-\frac{2f\rho ^{\prime 3}}{z}-\frac{\rho ^{\prime 2}}{%
\rho }-\left( \frac{2}{z}-\frac{f^{\prime }}{2f}\right) \rho ^{\prime }-%
\frac{1}{\rho f}=0\text{.}  \label{EOMC1}
\end{equation}%
If there are several extremal surfaces, the entropy functional is
multi-valued and the HEE is given by the minimal value.

\subsection{HEE and HSC}

\subsubsection{Large R expansion}

In Ref. \cite{Liu1202}, it has been pointed out that for $n=2$, the RT
surface only can be disk-like. While for $n>2$, the topology can be changed
from disk to cylinder as R increases, see Figure 6 in \cite{Liu1202} for a
schematic. In particular, we stress that for $n>2$ only a cylinder-like RT
surface can appear at large R.

Usually, it is not possible to solve the EOM for RT surfaces analytically.
But when R is large, the analytical solutions have been found in \cite%
{Liu1309} by a matching procedure. Let's briefly review the strategy. In the
UV, one can find that $\rho (z)$ can be expanded in 1/R as
\begin{equation}
\rho (z)=R-\frac{\rho _{1}(z)}{R}-\frac{\rho _{3}(z)}{R^{3}}+\cdots -\frac{%
\hat{\rho}(z)}{R^{v}}+\cdots ,  \label{roUV}
\end{equation}%
where the last term denotes the leading term of those that are not odd
powers of 1/R. We will focus on $n>2$ at first. For our aim, it is enough to
keep only the pending function $\rho _{1}(z)$. In terms of the EOM (\ref%
{EOMC1}), $\rho _{1}(z)$ should satisfy
\begin{equation}
\frac{z^{2}}{\sqrt{f}}(\frac{\sqrt{f}}{z^{2}}\rho _{1}^{\prime })^{\prime
}=s_{1},  \label{ro1}
\end{equation}%
where the source is $s_{1}=-1/f$. Equation (\ref{ro1}) has the solution%
\begin{equation}
\rho _{1}(z)=\int_{0}^{z}du\frac{u^{2}}{\sqrt{f(u)}}(b_{1}+\int_{u}^{\infty
}dv\frac{1}{v^{2}\sqrt{f(v)}}),  \label{ro12}
\end{equation}%
where $b_{1}$ is an integration constant. In the IR, the EOM (\ref{EOMC1})
implies that $\rho (z)$ has the large $z$ expansion%
\begin{equation}
\rho (z)=\rho _{0}+\frac{2z^{2-n}}{\rho _{0}a(n-2)(n+2)}+\cdots ,\quad n>2.
\label{roIR}
\end{equation}%
This is a cylinder-like solution. The constant $\rho _{0}$ means the radius
of the cylinder in the IR. To match the UV and IR, the solution (\ref{ro12})
is pulled to a sufficiently large $z$ so that Eq. (\ref{fzn}) holds. Then
one has%
\begin{equation}
\rho _{1}(z)=\frac{b_{1}}{\sqrt{a}}\frac{z^{3-n/2}}{3-n/2}(1+\cdots )+\frac{%
2z^{2-n}}{(2-n)(2+n)a}(1+\cdots ).  \label{ro13}
\end{equation}%
After inserting Eq. (\ref{ro13}) into the UV expansion (\ref{roUV}), it can
be matched to the IR expansion (\ref{roIR}), which leads to
\begin{equation}
b_{1}=0,\quad \rho _{0}=R.  \label{match}
\end{equation}%
Now one has a large-R solution applied to the total region of $z$.
Calculating Eq. (\ref{Area1}) with this solution, one can obtain
\begin{equation}
\mathcal{S}\sim R\int_{0}^{\infty }dz\frac{1}{z^{2}\sqrt{f(z)}}-\frac{1}{R}%
\int_{0}^{\infty }dz\frac{\sqrt{f(z)}}{2z^{2}}\rho _{1}^{\prime }(z)^{2}+%
\mathcal{O}(\frac{1}{R^{3}}),
\end{equation}%
where $\rho _{1}(z)$ is given by Eq. (\ref{ro13}) and Eq. (\ref{match}). One
can see that the TEE is zero.

The matching procedure for $n=2$ is similar. The final solution can be
written as\footnote{%
In the parentheses, we have neglected an exponential term$\sim \exp
(-2aR^{2})$. Such non-analytic term would be interesting in itself\ but it
is not important for us.}%
\begin{equation}
\rho (z)=R-\frac{1}{2\sqrt{a}}(\frac{b_{1}}{R}+\cdots )z^{2}-\frac{1}{2aR}%
\log {z}+\cdots ,  \label{ron2}
\end{equation}%
where $b_{1}=0$. Plugging Eq. (\ref{ron2}) into the entropy functional (\ref%
{Area1}) leads to%
\begin{equation}
\mathcal{S}\sim R\int_{0}^{\infty }dz\frac{1}{z^{2}\sqrt{f(z)}}+\frac{1}{R}%
\int_{0}^{\infty }dz\frac{f(z)-4az^{2}\log {z}}{8a^{2}z^{4}\sqrt{f(z)}}+%
\mathcal{O}(\frac{1}{R^{3}}),
\end{equation}%
where the TEE is still zero.

Now we will calculate the HSC. Consider the volume of the space enclosed by
the RT surface. The HSC can be defined by \cite{Alishahiha1509}:
\begin{equation}
\mathcal{C}=\frac{\mathrm{Volume}\left( \Sigma \right) }{8\pi LG_{N}},
\label{C1}
\end{equation}%
where $L$ is supposed to be the AdS radius. It should be noted that the
suitable length scale for a gapped phase might not be determined by the AdS
radius alone. In this paper, we will not explore this issue since we focus
on comparing the area of RT surfaces and the RT volume. Moreover, for the
similar reason, we don't care about other proposals for the holographic
complexity.

From the metric (\ref{metriczn}) with the constant time, one can obtain the
complexity functional:%
\begin{equation}
\mathcal{C}=\frac{L^{2}}{4G_{N}}\int_{0}^{\infty }dz\frac{1}{z^{3}\sqrt{f(z)}%
}\int_{0}^{\rho (z)}d\rho \rho .  \label{C2}
\end{equation}%
Using the large-R expansion of $\rho (z)$, we can expand Eq. (\ref{C2}) to%
\begin{equation}
\mathcal{C}\sim \left\{
\begin{array}{c}
R^{2}\int_{0}^{\infty }dz\frac{1}{z^{3}\sqrt{f(z)}}-\int_{0}^{\infty }dz%
\frac{2\rho _{1}(z)}{z^{3}\sqrt{f(z)}}+\mathcal{O}(\frac{1}{R^{2}}),\;n>2;
\\
R^{2}\int_{0}^{\infty }dz\frac{1}{z^{3}\sqrt{f(z)}}-\int_{0}^{\infty }dz%
\frac{\log (z)}{az^{3}\sqrt{f(z)}}+\mathcal{O}(\frac{1}{R^{2}}),\;n=2.%
\end{array}%
\right.
\end{equation}%
One can see that the R-linear term is vanishing in the HSC.

\subsubsection{Topology change}

We will study some toy-models with $n>2$. Following \cite{Liu1202}, we will
set the geometries (\ref{metriczn}) with a simple form $f(z)=1+z^{n}$.
We will focus on the topology change of RT surfaces. Since it happens at
finite R, we will resort to the numerical method. Usually, the EOM for RT
surfaces is integrated from IR to UV. So we need the IR boundary condition.
For the cylinder topology, it is nothing but Eq. (\ref{roIR}). For the disk
topology, it is convenient to replace $z$ with $\rho $ as one of the surface
coordinates. Accordingly, the EOM and IR boundary condition can be written by%
\begin{equation}
z^{\prime \prime }+\frac{z^{\prime 3}}{\rho f}+\left( \frac{2}{z}-\frac{%
f^{\prime }}{2f}\right) z^{\prime 2}+\frac{z^{\prime }}{\rho }+\frac{2f}{z}%
=0,
\end{equation}%
\begin{equation}
z(\rho )=z_{\ast }-\frac{f\left( z_{\ast }\right) }{2z_{\ast }}\rho
^{2}+\cdots ,
\end{equation}%
where $z_{\ast }$ is the radial location at the top of the disk. Using the
IR boundary conditions, we can numerically depict the RT surface and in turn
obtain the HEE and HSC. To exhibit the features of HEE and HSC clearly, it
is convenient to subtract the divergent terms that depend on the UV cutoff $%
\epsilon $ of $z$. Inserting the RT surface for the pure AdS spacetime $\rho
(z)=\sqrt{R^{2}-z^{2}}$ into Eq. (\ref{Area1}) and Eq. (\ref{C2}), one can
isolate the divergent terms of the HEE and HSC%
\begin{equation}
\mathcal{S}_{\mathrm{div}}=\frac{\pi L^{2}}{2G_{N}}\frac{R}{\epsilon },\;%
\mathcal{C}_{\mathrm{div}}=\frac{L^{2}}{8G_{N}}\left( \frac{R^{2}}{2\epsilon
^{2}}+\log \epsilon \right) .  \label{SCdiv}
\end{equation}%
Then we define the regularized quantities%
\begin{equation}
S=\mathcal{S-S}_{\mathrm{div}},\;C=\mathcal{C-C}_{\mathrm{div}}.
\end{equation}%
For convenience, we further rescale them by%
\begin{equation}
\bar{S}=S/\left\vert S_{0}\right\vert ,\;\bar{C}=C/\left\vert
C_{0}\right\vert .
\end{equation}%
where $S_{0}$ and $C_{0}$ are certain constants. Their concrete values are
not important for the relevant physics. To be clear, we define them as
follows: $S_{0}$ is the critical value of $S$ when the RT surface changes
the topology and $C_{0}$ is the value of $C$ when the branch with disk
topology meets the branch with cylinder topology.

\begin{figure}[th]
\centerline{
\includegraphics[width=.8\textwidth]{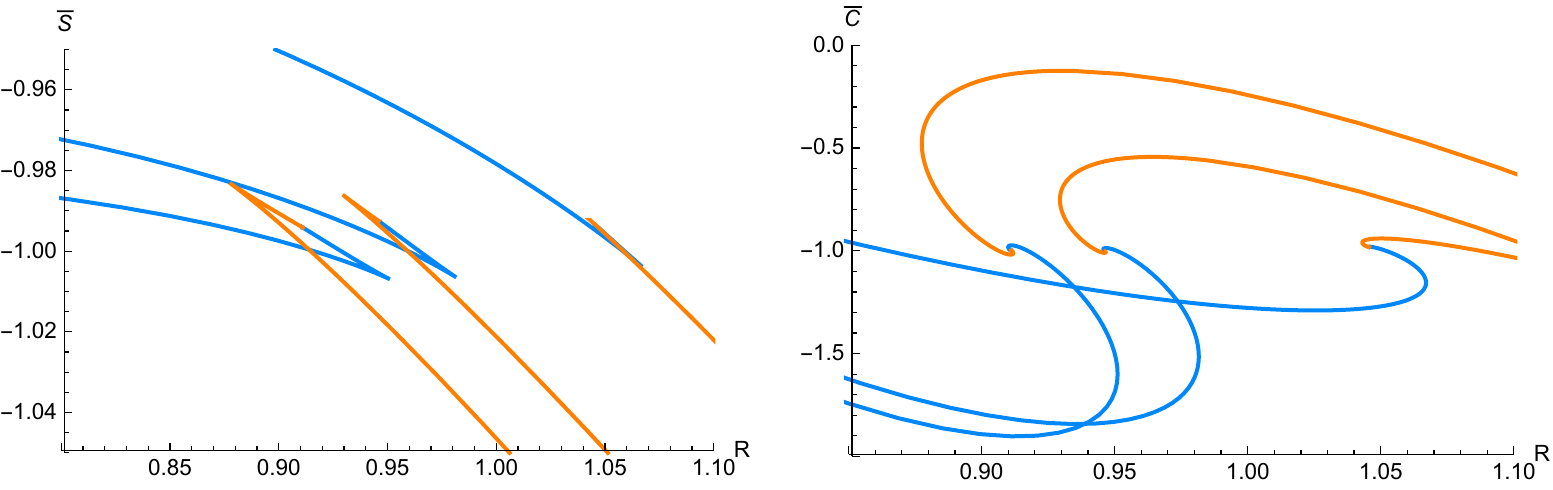}}
\caption{$\bar{S}$ (left) and $\bar{C}$ (right) as the functions of R for
the gapped geometries with singular IR. In each pannel, there are three
curves, which correspond to the different parameters $n=3.5$, $4.5$, $5.5$.
As $n$ decreases, the multi-valued regions gradually shrink. Each curve
includes two branches which have the disk topology (blue) and cylinder
topology (orange), respectively. This color scheme is used in all the
figures of the main text.}
\label{FIGIR}
\end{figure}

\begin{figure}[th]
\centerline{
\includegraphics[width=.8\textwidth]{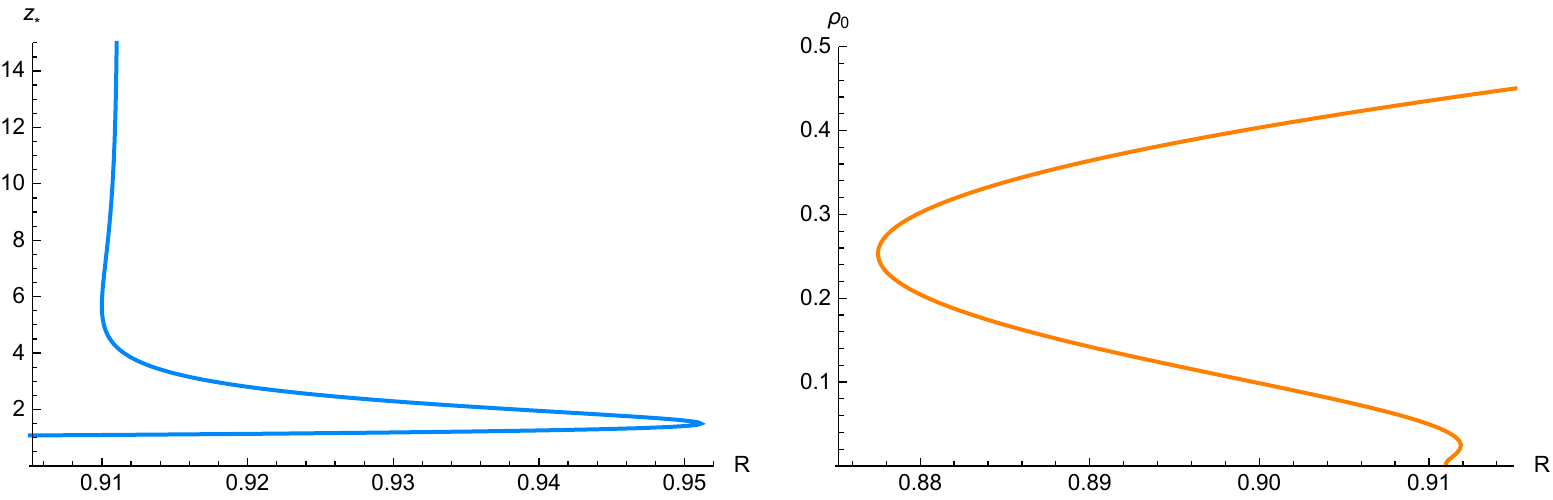}}
\caption{$z_{\ast }$ (left) and $\protect\rho _{0}$ (right) as the functions
of R for the gapped geometries with singular IR and $n=5.5$.}
\label{FIGRzs}
\end{figure}

We plot $\bar{S}$ and $\bar{C}$ in Figure \ref{FIGIR}. Some remarks are in
order. First, both entropy and complexity functionals are the multi-valued
functions of R near the critical radius at which the RT surface changes the
topology. Second, as $n$ decreases, their multi-valued regions gradually
shrink. For the small $n$, the complexity functional is still obviously
multi-valued but the entropy functional becomes approximately single-valued%
\footnote{%
When $n$ is close to $2$, we find that the multi-valued regions of $\bar{S}$
and $\bar{C}$ are difficult to be identified numerically. However, we
suspect that they would not disappear completely until $n$ is equal to 2, at
which the RT surface only has the disk topology.}. Third, the shapes of two
multi-valued regions are different: one is the classic `swallowtail' and the
other is not. Accordingly, the HEE (the minimal value of the entropy
functional) is continuous but its first derivative is not. On the contrary,
the HSC is discontinuous. In particular, the complexity functional at large $%
n$ exhibits a novel `double-S' behavior, while the entropy functional does
not have an obvious counterpart to the small `S'. Note that the existence of
small `S' region can be evidenced in the functions $z_{\ast }(R)$ and $\rho
_{0}(R)$, see Figure \ref{FIGRzs}. These results indicate that the HSC can
perceive the topology change of the RT surface. Compared to the HEE, it
reacts differently and can be more sensitive.

\bigskip

\section{AdS Soliton}

Consider the $\mathcal{N}=4$ super Yang-Mills theory in four dimensions,
with one compactified spatial direction $\phi $. Suppose that $\phi $ is
anti-periodic, which indicates massive fermions. At low energies, this
theory is reduced to the three-dimensional gauge theory, associated with the
confinement, mass gap and finite correlation length. By holography, it is
dual to the AdS$_{5}$ soliton in IIB string theory \cite{Witten9803}. The
AdS soliton is a gapped geometry since its compact dimension shrinks to zero
at some finite value of AdS radius, indicating an IR fixed point of the
field theory at finite energy scale.

\subsection{RT surface}

We will study the soliton geometry%
\begin{equation}
ds^{2}=\frac{{L^{2}}}{z^{2}}\left[ \frac{dz^{2}}{h(z)}-dt^{2}+d\rho
^{2}+\rho ^{2}d\theta ^{2}+h(z)d\phi ^{2}\right] +L^{2}d\Omega _{5}^{2},
\end{equation}%
where%
\begin{equation}
h(z)=1-(\frac{z}{z_{0}})^{4},
\end{equation}%
and $z_{0}$ is relevant to the period of $\phi $. Using the induced metric
for the RT surface%
\begin{equation}
ds_{\mathrm{ind}}^{2}=\frac{{L^{2}}}{z^{2}}\left[ \left( \frac{1}{h}+\rho
^{\prime 2}\right) dz^{2}+\rho ^{2}d\theta ^{2}+hd\phi ^{2}\right]
+L^{2}d\Omega _{5}^{2},
\end{equation}%
the entropy functional can be expressed as%
\begin{equation}
\mathcal{S}=\frac{\pi \Omega _{5}L_{\phi }L^{8}}{2G_{N}}\int_{0}^{z_{0}}dz%
\frac{\rho }{z^{3}}\sqrt{1+h\rho ^{\prime 2}},  \label{SsolitonD}
\end{equation}%
where $\Omega _{5}$ is the unit volume of five-dimensional sphere, $L_{\phi
} $ is the period of $\phi $, and $G_{N}$ is the ten-dimensional Newton
constant. Variation of the entropy functional (\ref{SsolitonD}) with respect
to $\rho (z)$ generates the EOM for the RT surface%
\begin{equation}
\frac{d}{dz}\left( \frac{h\rho \rho ^{\prime }}{z^{3}\sqrt{1+h\rho ^{\prime
2}}}\right) =\frac{1}{z^{3}}\sqrt{1+h\rho ^{\prime 2}}.  \label{EOMsoliton}
\end{equation}

\subsection{HEE and HSC}

\subsubsection{Large R expansion}

In \cite{Pakman0805}, it has been found that the RT surface is cylinder-like
as R is large. Also, it has been noticed that the solution of the EOM (\ref%
{EOMsoliton}) at large $\rho _{0}$ has the form%
\begin{equation}
\rho (z)=\rho _{0}+\frac{1}{\rho _{0}}\rho _{1}(z)+\cdots ,
\label{rozsoliton}
\end{equation}%
where $\rho _{1}(z)$ should be vanishing in the IR ($z\rightarrow z_{0}$).
Substituting this ansatz into (\ref{EOMsoliton}), one can derive%
\begin{equation}
\rho _{1}(z)=c_{2}+\frac{1-2c_{1}}{8}\log (1-z^{2})-\frac{1+2c_{1}}{8}\log
(1+z^{2}),  \label{ro1soliton}
\end{equation}%
where we have set $z_{0}=1$ for convenience. Noticing that the EOM (\ref%
{EOMsoliton}) near $z=1$ has a solution
\begin{equation}
\rho =\rho _{0}+\frac{1-z}{4\rho _{0}}+\cdots ,  \label{bcsoliton1}
\end{equation}%
the integral constants in (\ref{ro1soliton}) can be determined:%
\begin{equation}
c_{1}=\frac{1}{2},\;c_{2}=\frac{\log 2}{4}.
\end{equation}%
Furthermore, we know $\rho (0)=R$, which imposes%
\begin{equation}
\rho _{0}+\frac{1}{4\rho _{0}}\log (2)=R\text{.}
\end{equation}%
At large $\rho _{0}$, we have $\rho _{0}\simeq R$. Collecting all these
results together, one can recast Eq. (\ref{rozsoliton}) as the solution at
large R:%
\begin{equation}
\rho (z)=R+\frac{1}{4R}\log (\frac{1}{1+z^{2}})+\cdots .  \label{rozsoliton2}
\end{equation}%
Then Eq. (\ref{SsolitonD}) can be expanded as%
\begin{equation}
\mathcal{S}\sim R\int_{0}^{1}dz\frac{1}{z^{3}}+\frac{1}{R}\int_{0}^{1}dz%
\frac{1}{8z^{3}}\left[ \frac{z^{2}h}{(1+z^{2})^{2}}+2\log (\frac{1}{1+z^{2}})%
\right] +\mathcal{O}(\frac{1}{R^{3}}),  \label{Ssoliton2}
\end{equation}%
where the TEE is vanishing. This is the essential result in \cite{Pakman0805}%
. Since one does not expect any long-range order in the ground state of
(2+1) dimensional QCD, this result has been viewed as a consistency check on
the HEE at that time.

Now we turn to study the HSC, which can be determined by%
\begin{equation}
\mathcal{C=}\frac{\Omega _{5}L_{\phi }L^{8}}{4G_{N}}\int_{0}^{1}dz\frac{1}{%
z^{4}}\int_{0}^{\rho (z)}d\rho \rho .  \label{CsolitonD}
\end{equation}%
Using the large-R expansion (\ref{rozsoliton2}), we obtain%
\begin{equation}
\mathcal{C}\sim R^{2}\int_{0}^{1}dz\frac{1}{z^{4}}+\int_{0}^{1}dz\frac{1}{%
2z^{4}}\log (\frac{1}{1+z^{2}})+\mathcal{O}(\frac{1}{R^{2}}).
\end{equation}%
There is no R-linear term.

\subsubsection{Topology change}

The IR boundary condition (\ref{bcsoliton1}) can be used to solve
numerically the RT surface with cylinder topology. For disk topology, we
will look for the profile $z(\rho )$ of the RT surface, instead of $\rho (z)$%
. Rewrite the EOM (\ref{EOMsoliton}) by $z(\rho )$, from which the IR
boundary condition can be read:%
\begin{equation}
z(\rho )=z_{\ast }+\frac{z_{\ast }h^{\prime }(z_{\ast })-6h(z_{\ast })}{%
8z_{\ast }}\rho ^{2}+\cdots .
\end{equation}%
We also need the UV divergent terms of the HEE and HSC. In the UV $\left(
z\rightarrow 0\right) $, the bulk geometry approaches AdS$_{5}\times S^{5}$.
The EOM (\ref{EOMsoliton}) has the solution%
\begin{equation}
\rho (z)=R-\frac{z^{2}}{4R}+\frac{z^{4}\log z}{32R^{3}}+\cdots .
\end{equation}%
Substituting it into Eq. (\ref{SsolitonD}) and Eq. (\ref{CsolitonD}), the
divergent terms can be extracted:%
\begin{equation}
\mathcal{S}_{\mathrm{div}}=\frac{\pi \Omega _{5}L_{\phi }L^{8}}{4G_{N}}%
\left( \frac{R}{\epsilon ^{2}}+\frac{\log \epsilon }{4R}\right) ,\;\mathcal{C%
}_{\mathrm{div}}=\frac{\Omega _{5}L_{\phi }L^{8}}{4G_{N}}\left( \frac{R^{2}}{%
3\epsilon ^{3}}-\frac{1}{2\epsilon }\right) .
\end{equation}

\begin{figure}[th]
\centerline{
\includegraphics[width=.8\textwidth]{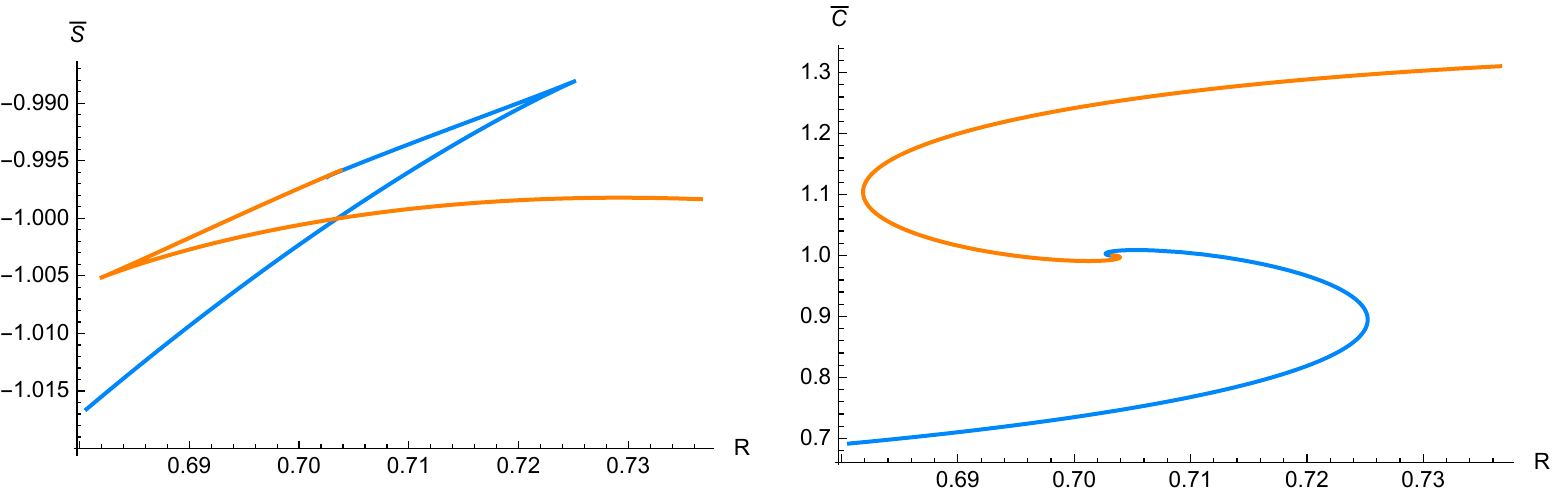}}
\caption{$\bar{S}$ (left) and $\bar{C}$ (right) as the functions of R for
the AdS soliton.}
\label{FIGsoliton}
\end{figure}

\begin{figure}[th]
\centerline{
\includegraphics[width=.8\textwidth]{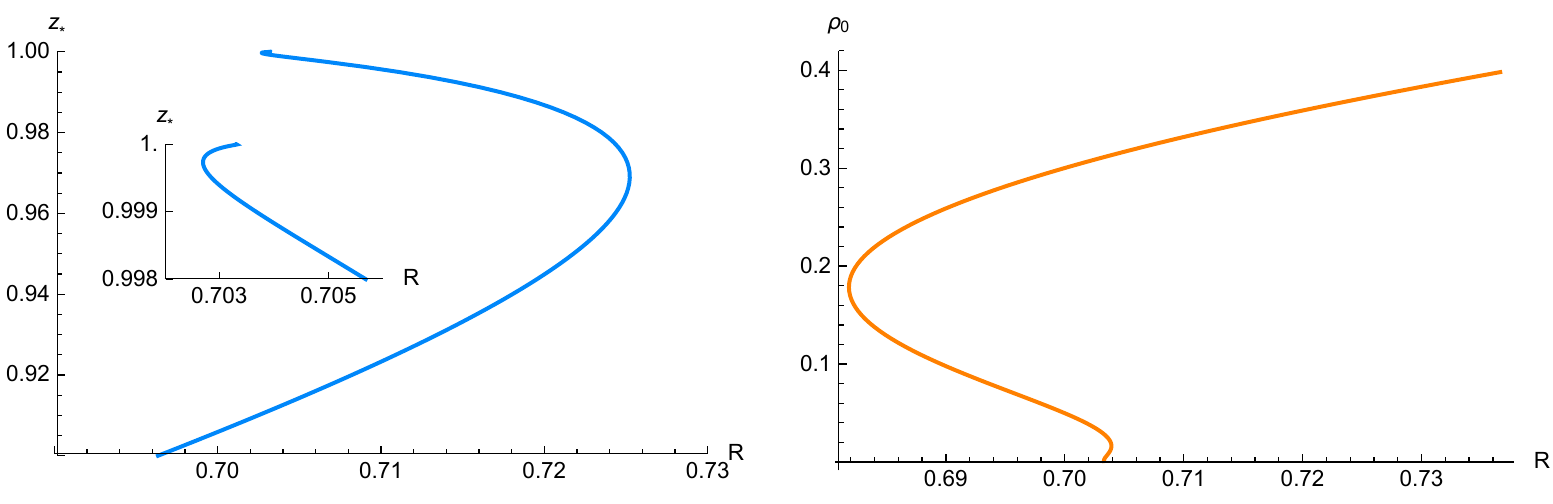}}
\caption{$z_{\ast }$ (left) and $\protect\rho _{0}$ (right) as the functions
of R for the AdS soliton.}
\label{FIGsoliton2}
\end{figure}

In the study of HEE \cite{Pakman0805}, it has been shown that there is a
`swallowtail' phase transition associated with the topology change of the RT
surface. In the left panel of Figure \ref{FIGsoliton}, we recover such a
behavior. From the right panel in Figure \ref{FIGsoliton}, we have found
that the HSC can probe the topology change in a different way: the
complexity functional behaves as the `double-S' instead of the
`swallowtail'. The existence of the `double-S' can be understood by the
double and triple valued regions of the functions $z_{\ast }(R)$ and $\rho
_{0}(R)$, see Figure \ref{FIGsoliton2}.

It should be noted that our figure for the HEE is not exactly same as Figure
6 in \cite{Pakman0805}. This is because their HEE subtracts a term $\sim
\left( \log R\right) /R$ besides the divergent terms. Moreover, the two
functions $z_{\ast }(R)$ and $\rho _{0}(R)$ have been plotted in Figure 4
and Figure 5 of \cite{Pakman0805}, respectively. Although the triple-valued
region has not been noticed in their Figure 4, it is barely visible in their
Figure 5.

\section{Soft-wall model}

The hard-wall and soft-wall models are both the bottom-up approaches to the
AdS/QCD duality. A basic feature of hard-wall models is the existence of an
IR brane at which the warped dimension abruptly ends. The hard IR wall
breaks the conformal symmetry and provides the simplest realization of the
confinement \cite{Polchinski2002}. In the soft-wall model, there is a
smoothing of the IR wall by invoking a dilaton field, which correctly
produces the Regge behavior of highly excited mesons \cite{Karch0602}.
Consider the power-law behavior of the dilaton $\phi (z)=(\mu z)^{\nu }$,
where $\mu $ is the energy scale and $z$ is the conformal coordinate. By
analyzing the eigenfunctions of bulk fields with such power-law dilaton, the
Kaluza-Klein mass spectrum with large $n$ can be given by $m_{n}^{2}\sim \mu
^{2}n^{2-2/\nu }$ \cite{Batell0808}. For $\nu <1$, the spectrum is gapless.
For $\nu =1$, the spectrum becomes gapped but is continuous above the gap.
For $\nu >1$, the spectrum is gapped and discrete.

\subsection{RT surface}

Suppose that the soft-wall geometry is described by%
\begin{equation}
ds^{2}=\frac{L^{2}a(z)}{z^{2}}(dz^{2}-dt^{2}+d\rho ^{2}+\rho ^{2}d\theta
^{2}).  \label{metricSW}
\end{equation}%
In the IR $\left( \mu z\gg 1\right) $ and UV $\left( \mu z\ll 1\right) $,
the warp factor are%
\begin{equation}
a_{\mathrm{IR}}(z)=e^{-(\mu z)^{\nu }},\;a_{\mathrm{UV}}(z)=1.
\end{equation}%
Following \cite{Parnachev1504}, we will be interested in an explicit example%
\begin{equation}
a(z)=\frac{1}{\cosh (\sqrt{(\mu z)^{2\nu }+1}-1},  \label{az}
\end{equation}%
which satisfies both boundary conditions. In Ref. \cite{Batell0801}, the
dynamical soft-wall model with the warp factor $a(z)=e^{-(\mu z)^{\nu }}$
has been constructed. Now we are expecting that the warp factor (\ref{az})
could be produced dynamically by a similar method. Note that the function (%
\ref{az}) is flatter than $e^{-(\mu z)^{\nu }}$ in the UV. This is helpful
to distinguish two space regions, namely Part (I) and Part (II), which will
be introduced below. Hereafter, we set $\mu =1$ for convenience.

From Eq. (\ref{metricSW}), we write down the induced metric for the RT
surface%
\begin{equation}
ds_{\mathrm{ind}}^{2}=\frac{L^{2}a(z)}{z^{2}}\left[ \left( 1+\rho ^{\prime
2}\right) dz^{2}+\rho ^{2}d\theta ^{2}\right] ,
\end{equation}%
which in turn gives the entropy functional%
\begin{equation}
\mathcal{S}=\frac{\pi L^{2}}{2G_{N}}\int_{0}^{\infty }dz\frac{a\rho }{z^{2}}%
\sqrt{1+\rho ^{\prime 2}}.  \label{SSW}
\end{equation}%
Taking the variation with respect to $\rho (z)$ derives the EOM for the RT
surface%
\begin{equation}
\frac{d}{dz}\left( \frac{a\rho }{z^{2}}\frac{\rho ^{\prime }}{\sqrt{1+\rho
^{\prime 2}}}\right) =\frac{a}{z^{2}}\sqrt{1+\rho ^{\prime 2}}.
\label{EOMSW}
\end{equation}

\subsection{HEE and HSC}

In \cite{Parnachev1504}, it has been shown that the RT surface only has the
disk topology for $\nu <2$. Thus, the range $1\leq \nu <2$ is desired for
the gapped geometry with a disk-like RT surface. Meanwhile, they proved that
the TEE is vanishing by an analytical method. The basic strategy is to split
the RT surface into three parts, see a schematic in Figure 1 of \cite%
{Parnachev1504}. Part (I) is the deep UV region with $\epsilon
<z<z_{c}^{(1)} $, where $z_{c}^{(1)}$ denotes the crossover scale at which
the warp factor changes from $a_{\mathrm{UV}}(z)$ to $a_{\mathrm{IR}}(z)$.
Part (II) is the intermediate region with $z_{c}^{(1)}<z<z_{c}^{(2)}$, and
part (III) is the deep IR region with $z>z_{c}^{(2)}$. Here $z_{c}^{(2)}$ is
chosen such that the surface area in part (II) is not exponentially
suppressed by the warp factor $a_{\mathrm{IR}}(z)$ but will be in part
(III). In part (I), the profile of the RT surface is known:%
\begin{equation}
\rho (z)=R-\frac{z^{2}}{2R}+\mathcal{O}(\frac{1}{R^{2}}).
\end{equation}%
In part (II), it was found that%
\begin{equation}
\rho (z)=\sqrt{2d_{1}}z_{0}^{1-\frac{\nu }{2}}-\frac{1}{\sqrt{2d_{1}}%
z_{0}^{1-\frac{\nu }{2}}}\frac{z^{2-\nu }}{\nu (2-\nu )}+\mathcal{O}(\frac{1%
}{z_{0}^{3-\frac{3\nu }{2}}}),  \label{part2}
\end{equation}%
where $d_{1}$ is an integration constant. The parameter $z_{0}$ is required
to obey $z\ll z_{0}$. By identifying $R=\sqrt{2d_{1}}z_{0}^{1-\frac{\nu }{2}%
} $, Eq. (\ref{part2}) is changed to%
\begin{equation}
\rho (z)=R-\frac{1}{R}\rho _{\mathrm{II}}(z)+\mathcal{O}(\frac{1}{R^{3}}),
\end{equation}%
where%
\begin{equation}
\rho _{\mathrm{II}}(z)=\frac{z^{2-\nu }}{\nu (2-\nu )}.
\end{equation}%
Thus, the RT surface in parts (I) and (II) can be taken together as
\begin{equation}
\rho (z)=R-\frac{\rho _{\mathrm{int}}(z)}{R}+\mathcal{O}(\frac{1}{R^{2}}),
\label{rozSW}
\end{equation}%
where $\rho _{\mathrm{int}}(z)$ is an interpolating function.

Inserting Eq. (\ref{rozSW}) into Eq. (\ref{SSW}), one can obtain%
\begin{equation}
\mathcal{S}=\mathcal{S}_{\mathrm{I}}+\mathcal{S}_{\mathrm{II}}+\mathcal{S}_{%
\mathrm{III}},
\end{equation}%
where%
\begin{equation}
\mathcal{S}_{\mathrm{I}}\sim R\int_{\epsilon }^{z_{c}^{(1)}}dz\frac{1}{z^{2}}%
,  \label{SI}
\end{equation}%
hence leading to a typical UV divergence, the interesting part is%
\begin{equation}
\mathcal{S}_{\mathrm{II}}\sim R\int_{z_{c}^{(1)}}^{z_{c}^{(2)}}dz\frac{a_{%
\mathrm{IR}}}{z^{2}}+\frac{1}{R}\int_{z_{c}^{(1)}}^{z_{c}^{(2)}}dz\frac{a_{%
\mathrm{IR}}}{z^{2}}(\frac{\rho _{\mathrm{II}}^{\prime 2}}{2}-\rho _{\mathrm{%
II}})+\mathcal{O}(\frac{1}{R^{2}}),  \label{SII}
\end{equation}%
and $\mathcal{S}_{\mathrm{III}}$ is exponentially suppressed by
construction. Thus, the TEE is equal to zero\footnote{%
In the previous models, the absence of TEE can be attributed to the
vanishing constant term in the profile $\rho (z)$. In the current model, we
emphasize that the expansion (\ref{rozSW}) is not applicable to Part (III).
This is because the RT surface has the disk topology and its profile in the
IR satisfies $\rho (z_{\ast })=0$ for any R. However, the TEE is still
vanishing since the entropy functional in Part (III) is exponentially
suppressed.}.

We can study the HSC in a similar way. The only thing to be careful is that $%
z_{c}^{(2)}$ should be large enough to exponentially suppress the RT volume
inside part (III) and $z_{0}$ (thereby R) should be so large that one has $%
z\ll z_{0}$ in part (II). Let's consider the complexity functional%
\begin{equation}
\mathcal{C}=\frac{L^{2}}{4G_{N}}\int_{0}^{\infty }dz\frac{a^{3/2}}{z^{3}}%
\int_{0}^{\rho (z)}d\rho \rho .
\end{equation}%
Using the large-R expansion (\ref{rozSW}), we read%
\begin{equation}
\mathcal{C}=\mathcal{C}_{\mathrm{I}}+\mathcal{C}_{\mathrm{II}}+\mathcal{C}_{%
\mathrm{III}},
\end{equation}%
where the UV divergence comes from%
\begin{equation}
\mathcal{C}_{\mathrm{I}}\sim \int_{\epsilon }^{z_{c}^{(1)}}dz\frac{%
R^{2}-z^{2}}{z^{3}},  \label{CI}
\end{equation}%
the interesting part is%
\begin{equation}
\mathcal{C}_{\mathrm{II}}\sim R^{2}\int_{z_{c}^{(1)}}^{z_{c}^{(2)}}dz\frac{%
a_{\mathrm{IR}}^{3/2}}{z^{3}}-2\int_{z_{c}^{(1)}}^{z_{c}^{(2)}}dz\frac{a_{%
\mathrm{IR}}^{3/2}}{z^{3}}\rho _{\mathrm{II}}+\mathcal{O}(\frac{1}{R}),
\end{equation}%
and $\mathcal{C}_{\mathrm{III}}$ is exponentially suppressed. One can see
that the R-linear term disappears.

\begin{figure}[th]
\centerline{
\includegraphics[width=.8\textwidth]{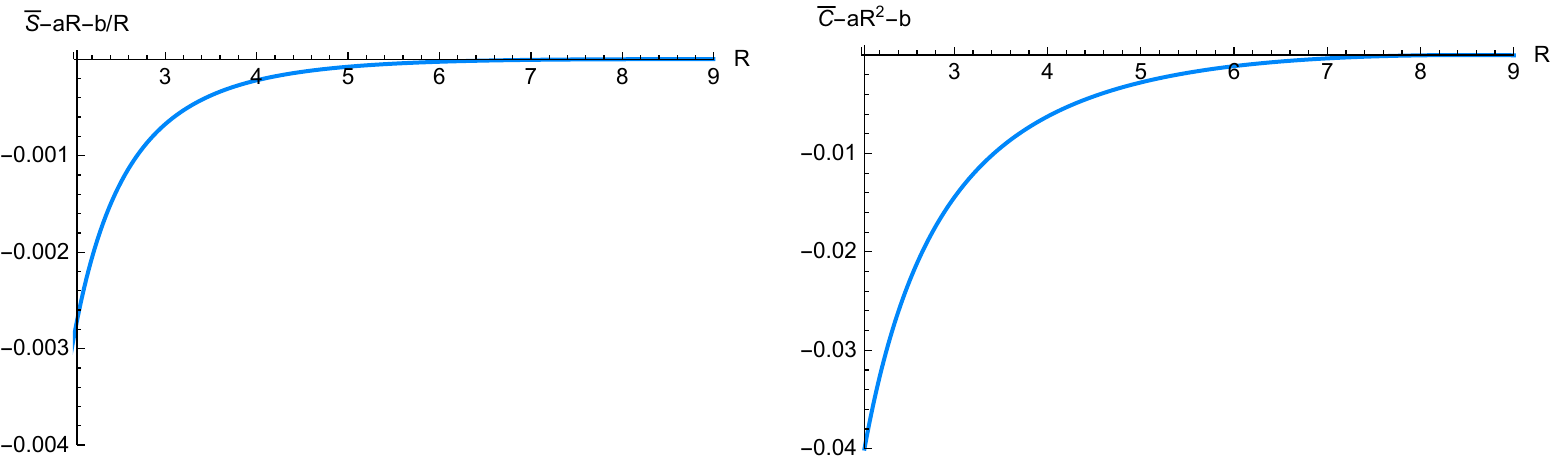}}
\caption{The large-R behavior of HEE (left) and HSC (right) in the soft-wall
model. Here $b_{1}$ and $b_{2}$ are the best fitting parameters.}
\label{FIGSW}
\end{figure}

In this paper, the absence of the R-linear term is mainly exhibited by
analytical methods. However, we have checked by numerical methods that this
is true for all the models. As an example, we will numerically calculate the
large-R behavior of HEE and HSC in the soft-wall model. Rewrite the EOM (\ref%
{EOMSW}) by the profile $z(\rho )$, which can be solved using the IR
boundary condition%
\begin{equation}
z(\rho )=z_{\ast }+\frac{z_{\ast }a^{\prime }(z_{\ast })-2a(z_{\ast })}{%
4z_{\ast }a(z_{\ast })}\rho ^{2}+\cdots .  \label{zrosw}
\end{equation}%
Then we\ use the functions $b_{1}R+b_{2}/R$ and $b_{1}R^{2}+b_{2}$ to fit
the large-R regions of $\bar{S}$ and $\bar{C}$, respectively\footnote{%
In the present model, the resealed factor $S_{0}$ and $C_{0}$ are defined by
the HEE and HSC at $R=2$. The index $\nu $ is fixed as 3/2. We have checked
that other values in the range $1\leq \nu <2$ do not change the result
qualitatively.}. The result is shown in Figure \ref{FIGSW}.

One might wonder whether the numerical parameters $b_{1}$ and $b_{2}$ match
their analytical expressions. For the previous two models, we find that they
match perfectly. Take the HEE of AdS soliton as an example. The error is
less than 0.01\% for $R\simeq 20$. For the soft-wall model, however, the
accuracy is limited since $z_{c}^{(1)}$ cannot be fixed accurately by its
definition and $a(z)$ and $\rho (z)$ are discontinuous between (I) and (II).
In fact, that is why we select to display the numerical result of the
soft-wall model. To be clear, let's compare the analytical expression of HEE
to the numerical result. The behavior of HSC is similar. From Eq. (\ref%
{SI}) and Eq. (\ref{SII}), we have%
\begin{eqnarray}
\mathcal{S}_{\mathrm{I}}+\mathcal{S}_{\mathrm{II}} &\sim &R\int_{\epsilon
}^{z_{c}^{(1)}}dz\frac{1}{z^{2}}+R\int_{z_{c}^{(1)}}^{z_{c}^{(2)}}dz\frac{a_{%
\mathrm{IR}}}{z^{2}}+\frac{1}{R}\int_{z_{c}^{(1)}}^{z_{c}^{(2)}}dz\frac{a_{%
\mathrm{IR}}}{z^{2}}(\frac{\rho _{\mathrm{II}}^{\prime 2}}{2}-\rho _{\mathrm{%
II}})+\mathcal{O}(\frac{1}{R^{2}})  \notag \\
&=&R\left( \frac{1}{\epsilon }+B_{1}\right) +\frac{1}{R}B_{2}+\mathcal{O}(%
\frac{1}{R^{2}}),
\end{eqnarray}%
where%
\begin{equation}
B_{1}=-\frac{1}{z_{c}^{(1)}}+\int_{z_{c}^{(1)}}^{z_{c}^{(2)}}dz\frac{a_{%
\mathrm{IR}}}{z^{2}},\;B_{2}=\int_{z_{c}^{(1)}}^{z_{c}^{(2)}}dz\frac{a_{%
\mathrm{IR}}}{z^{2}}(\frac{\rho _{\mathrm{II}}^{\prime 2}}{2}-\rho _{\mathrm{%
II}}).
\end{equation}%
We denote $\bar{b}_{1}$ and $\bar{b}_{2}$ as the rescaled parameters of $%
B_{1}$ and $B_{2}$, respectively. Following \cite{Parnachev1504}, we set $%
z_{c}^{(1)}=1$ and $z_{c}^{(2)}=\infty $. At $R\simeq 20$, we find $\bar{b}%
_{1}\simeq 1.4b_{1}$ and $\bar{b}_{2}\simeq 0.55b_{2}$. Thus, although the
analytical expression of HEE does not match the numerical result very well,
it does correctly reflect the fact that the TEE disappears.

\section{Holographic FQH effect}

The FQH effect is associated with a state of quantum fluids that is
dominated by strongly correlated electrons in high magnetic field. Due to
the strong interaction and unusual symmetry in essence that can be
implemented relatively easy in the holographic framework, it has been argued
that the FQH system is likely to be a profitable place to apply the AdS/CFT
correspondence \cite{Bayntun1008}. Of particular interest to us, the FQH
system is the prototype of topologically ordered medium that can be
experimentally realized \cite{Wen1990}. It has been found that the TEE in
the fermionic Laughlin states is related to the filling fraction \cite%
{Haque0609}. Thus, it would be interesting to study the HEE and HSC in
holographic FQH states. Early holographic researches based on either
bottom-up phenomenological approaches or top-down string/brane settings are
very fruitful, e.g., see \cite%
{KeskiVakkuri:2008eb,Davis:2008nv,Fujita:2009kw,Hikida:2009tp,Alanen:2009cn,Bergman:2010gm,Gubankova:2010rc,Jokela:2010nu,Fujita:2012fp,Melnikov:2012tb,Kristjansen:2012ny}%
.

Here we will focus on a recent bottom-up model \cite{Lippert1409}. It is an
Einstein-Maxwell-Axion-Dilaton theory with the SL(2,Z) symmetry, which not
only captures the modular duality among various FQH states\footnote{%
Note that some well-known experimental results, such as the duality relation
and the semi-circle law in the plateaux transitions, can be attributed to
the existence of the modular symmetry group which commutes with the
renormalization group flow \cite{Dolan98,Dolan99,Dolan02}.} but also has the
solution with the hard mass gap and correct Hall conductivity related to the
filling fraction. Apparently, the solution to model FQH states should have
both electric and magnetic charges. But ascribed to the SL(2,Z)
transformation, it can be generated from a solution with purely electric
charge. In Appendix A, we will review briefly how to numerically construct
the electric solution that is a RG flow from the UV fixed point to the
dilatonic scaling IR.

\subsection{RT surface}

Consider the solution in terms of the metric ansatz%
\begin{equation}
ds^{2}=e^{2A(r)}\left[ -f(r)dt^{2}+d\rho ^{2}+\rho ^{2}d\theta ^{2}\right] +%
\frac{dr^{2}}{f(r)}.
\end{equation}%
The numerical functions of $f(r)$ and\ $A(r)$ are plotted in Appendix A, see
Figure \ref{FIGFQH2}. The induced metric for the RT surface is%
\begin{equation}
ds_{\mathrm{ind}}^{2}=\left( e^{2A}\rho ^{\prime 2}+\frac{1}{f}\right)
dr^{2}+e^{2A}\rho ^{2}d\theta ^{2},
\end{equation}%
from which the entropy functional can be expressed as%
\begin{equation}
\mathcal{S}=\frac{\pi }{2G_{N}}\int_{0}^{\infty }dre^{A}\rho \sqrt{%
e^{2A}\rho ^{\prime 2}+\frac{1}{f}}.
\end{equation}%
Taking the variation with respect to $\rho (r)$, one can derive the EOM for
the RT surface%
\begin{equation}
\frac{d}{dr}\frac{e^{3A}\rho \rho ^{\prime }}{\sqrt{e^{2A}\rho ^{\prime 2}+%
\frac{1}{f}}}=e^{A}\sqrt{e^{2A}\rho ^{\prime 2}+\frac{1}{f}}.  \label{EOMQHE}
\end{equation}

\subsection{HEE and HSC}

\subsubsection{Large R expansion}

The IR metric is characterized by Eq. (\ref{IR1}) and Eq. (\ref{IR2}), with
some parameters. By coordinate transformations, it can be rewritten as
\begin{equation}
ds^{2}=\frac{L^{2}}{z^{2}}\left( \frac{1}{az^{n}}dz^{2}-dt^{2}+d\rho
^{2}+\rho ^{2}d\theta ^{2}\right) ,  \label{metric1}
\end{equation}%
where%
\begin{eqnarray}
n &=&\frac{4s}{s+1},  \label{a} \\
a &=&-\frac{256\gamma ^{4}\Lambda L^{2-n}}{c^{2}\left( n-4\right)
^{4}u\omega }\exp (nA_{\mathrm{IR}}-\frac{n}{n-4}\gamma \phi _{\mathrm{IR}}).
\label{aa}
\end{eqnarray}%
Note that the singular behavior of Eq. (\ref{aa}) at $n=4$ is forbidden due
to the definition (\ref{a}).

From Eq. (\ref{metric1}), the holographic FQH model can be viewed as a
concrete realization of the general geometries that studied in Sec. 2. To go
ahead, we need to specify the parameter $n$. Consider the allowed region of
parameters $s$ and $\gamma $. It is tiny since a sensible theory is required
to satisfy five constraints, such as the absence of naked singularities at
finite temperatures and the presence of mass gap, see Sec. 3.2 and Figure 2
in \cite{Lippert1409}. Then one can read that the minimum value of $n$ is 2
when $s=1$ \cite{Parnachev1504}. In terms of Sec. 2, we infer that the TEE
is vanishing and the HSC does not have the R-linear term. In other words,
the absence of the constant term in the HEE and the R-linear term in the HSC
depends on the geometry in the IR. Nevertheless, we have made double check
of the large-R behavior by numerical methods, for which the complete
background is needed.

\subsubsection{Topology change}

Following \cite{Lippert1409}, we choose the parameters%
\begin{equation}
\gamma =-0.85,\;s=1.2
\end{equation}%
which indicates the IR index $n=24/11>2$. Thus, the topology of the RT
surface should change from disk to cylinder as R increases. Using the IR
solutions (\ref{IR1})-(\ref{IR5}), two kinds of the IR boundary conditions
of the RT surface can be derived from the EOM (\ref{EOMQHE}) and its
variant. They are%
\begin{eqnarray}
r(\rho ) &=&r_{\ast }+\frac{1}{2}e^{2A(r_{\ast })}f(r_{\ast })A^{\prime
}(r_{\ast })\rho ^{2}+\cdots ,\;\text{disk,} \\
\rho (r) &=&\rho _{0}+\frac{v_{1}}{\rho _{0}}\left[ p\left( r-r_{0}\right) %
\right] ^{v_{2}}+\cdots ,\;\text{cylinder,}
\end{eqnarray}%
where%
\begin{equation}
v_{1}=-\frac{c^{2}u\omega \left( n-4\right) ^{2}}{128\left( n^{2}-4\right)
\gamma ^{4}\Lambda }\exp (-2A_{\mathrm{IR}}+\frac{n}{n-4}\gamma \phi _{%
\mathrm{IR}}),\;v_{2}=4\gamma ^{2}\frac{n-2}{\left( n-4\right) ^{2}}.
\end{equation}%
We need to write down the HSC%
\begin{equation}
\mathcal{C}=\frac{1}{4G_{N}L}\int_{0}^{\infty }dr\frac{e^{2A}}{\sqrt{f}}%
\int_{0}^{\rho (r)}d\rho \rho .
\end{equation}%
Using the solution in the UV%
\begin{equation}
r(\rho )=-\frac{L}{2}\log \frac{R^{2}-\rho ^{2}}{L^{2}},
\end{equation}%
we isolate the UV divergences in the HEE and HSC, which is
\begin{equation}
\mathcal{S}_{\mathrm{div}}=\frac{\pi L^{2}}{2G_{N}}\left( \frac{R}{L}e^{%
\frac{r_{\mathrm{inf}}}{L}}\right) ,\;\mathcal{C}_{\mathrm{div}}=\frac{L^{2}%
}{8G_{N}}\left[ \frac{1}{2}\left( \frac{R}{L}e^{\frac{r_{\mathrm{inf}}}{L}%
}\right) ^{2}-\frac{r_{\mathrm{inf}}}{L}\right] ,  \label{UV}
\end{equation}%
where $r_{\mathrm{inf}}$ is the UV cutoff. Note that the UV divergences (\ref%
{UV}) are the same as those in pure AdS. In fact, Eq. (\ref{UV}) can be
derived from Eq. (\ref{SCdiv}) by a coordinate transformation $r=-L\log
\frac{z}{L}$.

Collecting these results together and using the metric constructed in
Appendix A, we can numerically calculate the HEE and HSC, see Figure \ref%
{FIGFQH}. One can find that there is no obvious signal indicating the
topology change of the RT surface. This might be relevant to the fact that
the IR index $n$ is close to $2$, at which only the disk topology exists.
\begin{figure}[th]
\centerline{
\includegraphics[width=.8\textwidth]{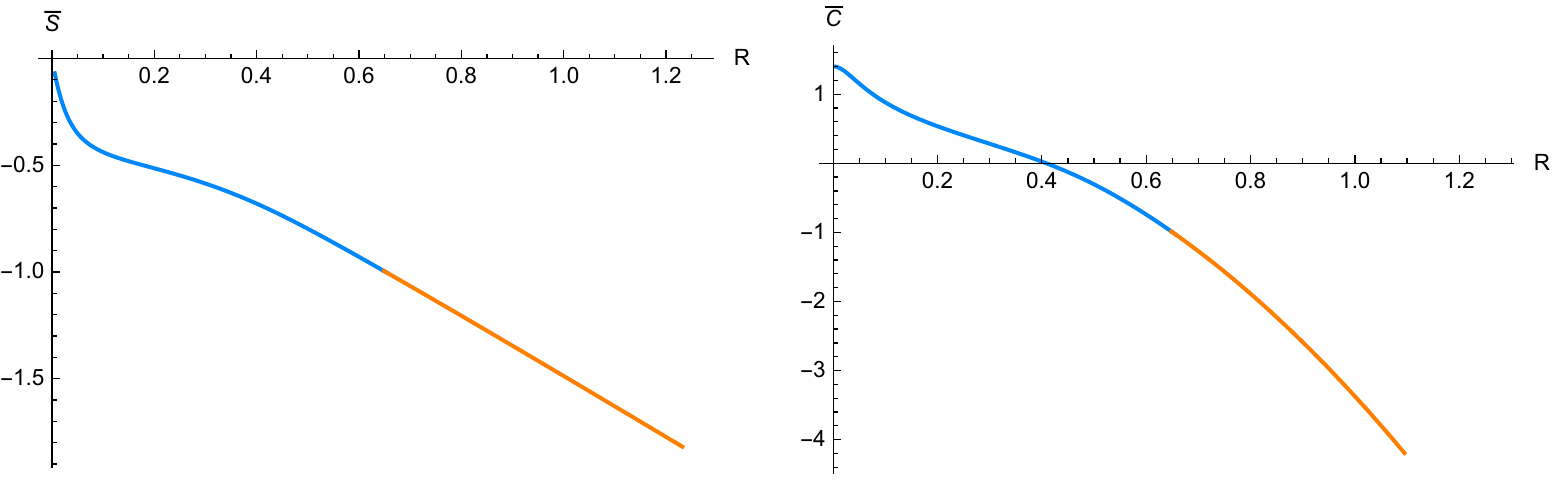}}
\caption{$\bar{S}$ (left) and $\bar{C}$ (right) as the functions of R for
the FQH model.}
\label{FIGFQH}
\end{figure}

\section{Gauss-Bonnet curvature}

TEE is the logarithm of the total quantum dimension. For FQH states, it can
be calculated by the Chern-Simons theory \cite{Dong0802}. For theories
compactified on a higher genus surface, the TEE is related to the degeneracy
of the ground state. In \cite{Parnachev1504}, a similar relation has been
exhibited in holography, for which both the TEE and ground state degeneracy
are produced by the GB curvature in the bulk. This holographic mechanism
requires two conditions: i) the RT surface is disk-like; ii) the TEE is
suppressed, except the GB contribution. Both of them have been realized in
the soft-wall model that we studied in Sec. 4. Here we will study the GB
contribution to the HSC in the same model.

It is implicitly assumed that the previous definition of the HSC (\ref{C1})
is applicable only to the Einstein gravity. Motivated by the Wald entropy
\cite{Wald93}, the HSC of a higher-derivative gravity theory was proposed
\cite{Alishahiha1509}:%
\begin{equation}
\mathcal{C=}\frac{1}{L}\int_{\mathcal{B}}\sqrt{\sigma }d^{3}xE^{\mu \nu
\lambda \rho }\epsilon _{\mu \nu }\epsilon _{\lambda \rho },\;E^{\mu \nu
\lambda \rho }=\frac{\partial \mathcal{L}}{\partial R_{\mu \nu \lambda \rho }%
}.  \label{CHD0}
\end{equation}%
Here $\mathcal{L}$ is Lagrangian, $\mathcal{B}$ is the bulk space enclosed
by the RT surface, $\sigma $ is the determinant of the space, and $\epsilon
_{\mu \nu }$ should be certain binormal. As pointed out in \cite{Bueno1612},
however, such definition suffers from the arbitrariness of the foliation in $%
\mathcal{B}$. Nevertheless, it was speculated that the complexity functional
in high-derivative theories should involve contractions of the 4-rank tensor
$E^{\mu \nu \lambda \rho }$ and the geometric quantities characterizing $%
\mathcal{B}$. As a result, a general complexity functional involving $E^{\mu
\nu \lambda \rho }$ at most once was presented:%
\begin{equation}
\mathcal{C}\mathcal{=}\frac{1}{L}\int_{\mathcal{B}}\sqrt{\sigma }%
d^{3}xE^{\mu \nu \lambda \rho }\left[ \left( \beta _{1}n_{\mu }h_{\nu
\lambda }n_{\rho }+\beta _{2}h_{\mu \rho }h_{\nu \lambda }\right) +\beta _{3}%
\right] ,  \label{CHD}
\end{equation}%
where $h_{\mu \nu }$ and $n_{\mu }$ are the induced metric and the normal
vector of the space $\mathcal{B}$, respectively. The constants ($\beta
_{1},\beta _{2},\beta _{3}$) are constrained so that the total functional is
reduced to the volume functional for the Einstein gravity. Note that the
complexity of higher-derivative gravity theories based on both CA and CV
conjectures has been discussed in \cite{Alishahiha1702}. In the following,
we will study Eq. (\ref{CHD}).

For our aim, we split Eq. (\ref{CHD}) into%
\begin{equation}
\mathcal{C}\mathcal{=\mathcal{C}_{\mathrm{E}}+C}_{\mathrm{GB}},
\end{equation}%
where $\mathcal{\mathcal{C}_{\mathrm{E}}}$ depends only on the Einstein
gravity and $\mathcal{C}_{\mathrm{GB}}$ denotes the GB correction.
Importantly, in the 4-dimensional theory of gravity that we are concerning,
the GB curvature contributes only a topological term to the gravity action
and HEE \cite{Parnachev1504}. Therefore, neither the spacetime metric nor
the RT surface would be changed. Since we have shown in Sec. 4 that there is
no R-linear term in $\mathcal{\mathcal{C}_{\mathrm{E}}}$, we only need to
focus on $\mathcal{C}_{\mathrm{GB}}$.

Let's write down the GB part of the Lagrangian%
\begin{equation}
\mathcal{L}_{\mathrm{GB}}=\frac{\alpha }{16\pi G_{N}}(R_{\mu \nu \lambda
\rho }R^{\mu \nu \lambda \rho }+R^{2}-4R_{\mu \nu }R^{\mu \nu }),
\end{equation}%
where $\alpha $ is the GB coupling. Accordingly, the 4-rank tensor is%
\begin{eqnarray}
E_{\mathrm{GB}}^{\mu \nu \lambda \rho } &=&\frac{\alpha }{16\pi G_{N}}{\Big [%
}2R^{\mu \nu \lambda \rho }+\left( g^{\mu \lambda }g^{\nu \rho }-g^{\mu \rho
}g^{\nu \lambda }\right) R  \notag \\
&&+2\left( g^{\nu \lambda }R^{\mu \rho }-g^{\mu \lambda }R^{\nu \rho
}+g^{\mu \rho }R^{\nu \lambda }-g^{\nu \rho }R^{\mu \lambda }\right) {\Big ].%
}  \label{Eabcd1}
\end{eqnarray}%
Then $\mathcal{C}_{\mathrm{GB}}$ can be written as%
\begin{eqnarray}
\mathcal{C}_{\mathrm{GB}} &\mathcal{=}&\mathcal{C}_{1}+\mathcal{C}_{2}, \\
\mathcal{C}_{1} &=&\frac{\beta _{1}}{L}\int_{\mathcal{B}}\sqrt{\sigma }%
d^{3}xE_{\mathrm{GB}}^{\mu \nu \lambda \rho }n_{\mu }h_{\nu \lambda }n_{\rho
},  \label{C11} \\
\mathcal{C}_{2} &=&\frac{\beta _{2}}{L}\int_{\mathcal{B}}\sqrt{\sigma }%
d^{3}xE_{\mathrm{GB}}^{\mu \nu \lambda \rho }h_{\mu \rho }h_{\nu \lambda }.
\label{C22}
\end{eqnarray}%
Using the metric (\ref{metricSW}) for the soft-wall model and the 4-rank
tensor (\ref{Eabcd1}), one can calculate Eq. (\ref{C11}) and Eq. (\ref{C22}):%
\begin{eqnarray}
\mathcal{C}_{1} &=&\frac{\alpha \beta _{1}}{8G_{N}}\int_{0}^{\infty }dz\chi
_{1}\frac{a^{3/2}}{z^{3}}\int_{0}^{\rho (z)}d\rho \rho ,  \label{C1GB} \\
\mathcal{C}_{2} &=&\frac{\alpha \beta _{2}}{8G_{N}}\int_{0}^{\infty }dz\chi
_{2}\frac{a^{3/2}}{z^{3}}\int_{0}^{\rho (z)}d\rho \rho ,  \label{C2GB}
\end{eqnarray}%
where%
\begin{eqnarray}
\chi _{1} &=&-\frac{6}{a}+\frac{2za^{\prime }}{a^{2}}+\frac{3z^{2}a^{\prime
2}}{2a^{3}}-\frac{2z^{2}a^{\prime \prime }}{a^{2}}, \\
\chi _{2} &=&\frac{12}{a}-\frac{8za^{\prime }}{a^{2}}+\frac{2z^{2}a^{\prime
\prime }}{a^{2}}.
\end{eqnarray}%
To carry these integrals, the profile $\rho (z)$ obtained in Sec. 4 can be
used. So let's use the large-R expansion (\ref{rozSW}) and rewrite Eq. (\ref%
{C1GB}) and Eq. (\ref{C2GB}) as
\begin{equation}
\mathcal{C}_{i}=\mathcal{C}_{i,\mathrm{I}}+\mathcal{C}_{i,\mathrm{II}}+%
\mathcal{C}_{i,\mathrm{III}},\;i=1,2,
\end{equation}%
where%
\begin{equation}
\mathcal{C}_{i,\mathrm{I}}\sim \int_{\epsilon }^{z_{c}^{(1)}}dz\frac{%
R^{2}-z^{2}}{z^{3}},
\end{equation}%
\begin{equation}
\mathcal{C}_{i,\mathrm{II}}\sim R^{2}\int_{z_{c}^{(1)}}^{z_{c}^{(2)}}dz\chi
_{i}\frac{a^{3/2}}{z^{3}}-2\int_{z_{c}^{(1)}}^{z_{c}^{(2)}}dz\chi _{i}\frac{%
a^{3/2}}{z^{3}}\rho _{1}+\mathcal{O}(\frac{1}{R}),
\end{equation}%
and $\mathcal{C}_{i,\mathrm{III}}$ is exponentially suppressed by a suitable
selection of $z_{c}^{(2)}$. One can see that there is no R-linear term in $%
\mathcal{C}_{\mathrm{GB}}$.

\section{Conclusion}

Using the gauge/gravity duality, we studied the complexity of the disk
subregion in various (2+1)-dimensional gapped systems. We compared the HSC
and the HEE from two aspects.

Firstly, we found that the R-linear term in the HSC is absent in the large-R
expansion. In particular, it disappears for the similar reason that the TEE
disappears in the HEE\footnote{%
For most models in this paper, their disappearance is originated from the
absence of the constant term in the large-R expansion of the profile $\rho
(z)$. For the soft-wall model, the origins not only include the absence of
the constant term of $\rho (z)$ in parts (I) and (II) but also the
exponential suppression of the HEE or the HSC in part (III).}. This simple
but interesting result suggests that there might be an underlying relation
between the HSC and the topological order. However, we further showed that
the GB curvature in the soft-wall model cannot produce the R-linear term in
the HSC. If the R-linear term in the HSC is really correlated to the TEE, it
would be needed to refine the present conjecture for the HSC in the GB
gravity. In Appendix B, we will make some speculations on what the expected
expression would look like.

Secondly, when the entanglement entropy probes the classic `swallowtail'
phase transition, the complexity is associated with a novel `double-S'
behavior. Our result supports to take the HSC as a good order parameter for
some phase transitions \cite{Mazhari1601,Roy1701,Wang1704}. However, neither
the HEE nor the HSC can obviously exhibit the topology change of the RT
surface in the holographic FQH model \cite{Lippert1409}. More sensitive
order parameter might be required\footnote{%
We have checked that the derivatives of the HEE and the HSC with respect to
R are still continuous in the FQH model. Thus, from the perspective of the
HEE or the HSC, the topology change is not simply a \textquotedblleft
first-order\textquotedblright\ nor \textquotedblleft
second-order\textquotedblright\ phase transition. Moreover, the renormalized
entanglement entropy that is made of the HEE and its derivative \cite%
{Liu1202,Liu1309} is not sensitive to the topology change, either.}.
Moreover, we have shown that the complexity can be more sensitive than the
entanglement entropy to probe the topology change of the RT surface\footnote{%
It means that (i) sometimes the multi-valued region of $\bar{S}$ is not as
obvious as that of $\bar{C}$, see Figure \ref{FIGIR}; (ii) the small `S'
region of $\bar{C}$ exhibits the details of the extremal surfaces near the
transition point, which has not the obvious counterpart in $\bar{S}$, see
Figure \ref{FIGIR}-Figure \ref{FIGsoliton2}; (iii) the HEE has a
discontinuous first derivative along R, while the HSC itself is
discontinuous, see Figure \ref{FIGIR} and Figure \ref{FIGsoliton}.}. Interestingly, a similar result also appeared in a recent work: by quantifying the complexity of
subregions via their purification, it was demonstrated that the complexity
can perceive some features insensitive to the entanglement entropy
\cite{Camargo1807}.

In the future, it may be worth studying the HSC through the tensor network.
Among others \cite{Zhou0709,Stanford1406,Miyaji1503,Caputa1703,Czech1706},
this direction is motivated by the following work. In \cite{Lin1203}, by
comparing the minimal surfaces in the AdS soliton and the MERA (multi-scale
entanglement renormalization ansatz) network \cite{MERA}, it was argued that
the IR fixed-point state is the product state, which is consistent with the
vanishing TEE. In \cite{Rene1710}, using the AdS$_{3}$/CFT$_{2}$ duality, it
was found that the subregion complexity exhibits a topological discontinuity
as the RT surface changes the configuration. Similar results have also been
obtained using the CFT and the random tensor network \cite{Hayden1601}. It
would be interesting to explore whether the vanishing R-linear term in the
HSC could be reinterpreted by the tensor network. Moreover, by using the
kinematic space as a bridge, it was pointed out recently that the HSC
of a pure state can be represented by the HEE alone. Even for the excited
states they considered, a part of the HSC can be determined by the HEE \cite%
{Abt1805}. This paves a way to clarify the relation between the HEE and the
HSC in the field theory. In particular, along this line, it would be
promising to identify what are the counterparts of the TEE and the
`swallowtail' in the HSC. Our results should be useful in this regard.

\section*{Acknowledgments}

We thank Ren\'{e} Meyer\ for explaining the FQH model carefully. We thank
Masamichi Miyaji and Run-Qiu\ Yang for helpful discussions on the
holographic complexity. We also thank Mikio Nakahara and Qian Niu for kindly
answering our questions on topological invariants. We were supported
partially by NSFC with grants No. 11675097 and No. 11675140.

\appendix

\section{Numerical solution of the FQH model}

We will study the Einstein-Maxwell-Axion-Dilation theory with the SL(2,Z)
invariance:%
\begin{eqnarray}
S &=&\frac{1}{16\pi G_{N}}\int d^{4}x\sqrt{-g}\left( L_{g}+L_{F}+V\right) ,
\\
L_{g} &=&R-\frac{1}{2}(\partial \phi )^{2}-\frac{1}{2}\frac{e^{-2\gamma \phi
}}{\gamma ^{2}}(\partial \tau _{1})^{2}, \\
L_{F} &=&-\frac{1}{4}\left( e^{\gamma \phi }F^{2}+\frac{\tau _{1}}{2}%
\epsilon ^{\mu \nu \rho \sigma }F_{\mu \nu }F_{\rho \sigma }\right) , \\
V &=&\sum_{m,n\in Z}^{\prime }\left( \frac{|m+n\tau |^{2}}{\tau _{2}}\right)
^{-s},
\end{eqnarray}%
where $\tau =\tau _{1}+i\tau _{2}$ and $\tau _{2}=e^{\gamma \phi }$. The
prime above $\sum $ stands for $m,n\neq 0$. With the ansatz for the metric%
\begin{equation}
ds^{2}=e^{2A(r)}\left[ -f(r)dt^{2}+dx^{2}+dy^{2}\right] +\frac{dr^{2}}{f(r)}
\end{equation}%
and the nonvanishing component of the gauge potential $A_{t}(r)$, one can
derive five field equations%
\begin{eqnarray}
\tau _{2}^{\prime 2}+4\gamma ^{2}\tau _{2}^{2}A^{\prime \prime }+\tau
_{1}^{\prime 2} &=&0,  \label{A16} \\
f^{\prime \prime }+3A^{\prime }f^{\prime }-e^{-4A}\frac{q^{2}}{\tau _{2}}
&=&0,  \label{A17} \\
\tau _{1}^{\prime \prime }+\left( 3A^{\prime }+\frac{f^{\prime }}{f}-2\frac{%
\tau _{2}^{\prime }}{\tau _{2}}\right) \tau _{1}^{\prime }+\frac{\gamma
^{2}\tau _{2}^{2}}{f}\frac{\partial V}{\partial \tau _{1}} &=&0,  \label{A18}
\\
\left( \log \tau _{2}\right) ^{\prime \prime }+\left( 3A^{\prime }+\frac{%
f^{\prime }}{f}\right) (\log \tau _{2})^{\prime }+\frac{\tau _{1}^{\prime 2}%
}{\tau _{2}^{2}}+\frac{\gamma ^{2}\tau _{2}}{f}\left( \frac{\partial V}{%
\partial \tau _{2}}+\frac{1}{2}e^{-4A}\frac{q^{2}}{\tau _{2}^{2}}\right)
&=&0,  \label{A19} \\
-\frac{1}{2}\left( \frac{\tau _{2}^{2}}{\gamma \tau _{2}}\right)
^{2}+6A^{\prime 2}+2A^{\prime }\frac{f^{\prime }}{f}-\frac{V}{f}+\frac{1}{2f}%
e^{-4A}\frac{q^{2}}{\tau _{2}^{2}} &=&0.  \label{A20}
\end{eqnarray}%
The potential can be expanded for large $\tau _{2}$ as
\begin{eqnarray}
V &=&2\zeta (2s)\tau _{2}^{s}+2\sqrt{\pi }\tau _{2}^{1-s}\frac{\Gamma (s-1/2)%
}{\Gamma (s)}\zeta (2s-1)  \notag \\
&&+\frac{2\pi ^{s}\sqrt{\tau _{2}}}{\Gamma (s)}\sum_{m,n\in Z}^{\prime }|%
\frac{m}{n}|^{s-1/2}K_{s-1/2}(2\pi \tau _{2}|mn|)e^{2i\pi mn\tau _{1}},
\end{eqnarray}%
where $K$ denotes the modified Bessel function of the second kind. In the
IR, it has the form $V=-2\Lambda \tau _{2}^{s}$, where $\Lambda =-\zeta (2s)$
and $s$ is a real parameter. It is found that there is an extremal scaling
solution in the IR \cite{Lippert1409,Charmousis1005}%
\begin{eqnarray}
A(r) &=&A_{\mathrm{IR}}+\frac{(\gamma -\delta )^{2}}{4}\log (p(r-r_{0})),
\label{IR1} \\
f(r) &=&\frac{-16\Lambda }{\omega uc^{2}}e^{-\delta \phi _{\mathrm{IR}%
}}(p(r-r_{0}))^{v},  \label{IR2} \\
\phi (r) &=&\phi _{\mathrm{IR}}+(\delta -\gamma )\log (p(r-r_{0})),
\label{IR3} \\
\tau _{1}(r) &=&a_{0},  \label{IR4} \\
A_{t}(r) &=&\frac{8}{\omega }\sqrt{\frac{v\Lambda }{u}}e^{-\frac{\gamma
+\delta }{2}\phi _{\mathrm{IR}}}(p(r-r_{0}))^{\frac{\omega }{4}},
\label{IR5}
\end{eqnarray}%
where $\omega =3\gamma ^{2}-\delta ^{2}-2\gamma \delta +4$, $u=\gamma
^{2}-\gamma \delta +2$, $v=-\delta ^{2}+\gamma \delta +2$, $\delta =-\gamma
s $, and
\begin{equation}
q^{2}=e^{-4A_{\mathrm{IR}}}\frac{-4\Lambda e^{(\gamma -\delta )\phi _{%
\mathrm{IR}}}(2+\gamma \delta -\delta ^{2})}{(2+\gamma ^{2}-\gamma \delta )}%
\frac{p^{2}}{c^{2}}.  \label{qfaiIR}
\end{equation}

Following \cite{Lippert1409}, we will explain how to construct a numerical
solution of field equations. First, we will reduce two independent
parameters: the parameter $\phi _{\mathrm{IR}}$ can be determined by Eq. (%
\ref{qfaiIR}) and one can infer $c=p$ from Eq. (\ref{A19}). Second, given
six parameters $\left( \gamma ,s,a_{0},p,q,r_{0}\right) $, we can use Eqs. (%
\ref{IR1})-(\ref{IR5}) as the IR boundary conditions to obtain a trial
solution, where the remained parameter $A_{\mathrm{IR}}$ is turned to
respect $6A^{\prime 2}f=6/L^{2}$ in the UV. This is required by Eq. (\ref%
{A20}). Here $L=\sqrt{6/V(\tau _{\ast },\bar{\tau}_{\ast })}$ is the AdS
radius and $\tau _{\ast }$ denotes certain UV fixed point. Third, the field
equations are invariant under two transformations%
\begin{eqnarray}
r &\rightarrow &\lambda _{1}r,\;f\rightarrow \lambda _{1}^{2}f,
\label{tran1} \\
e^{A} &\rightarrow &\lambda _{2}e^{A},\;q\rightarrow \lambda _{2}^{2}q,
\label{tran2}
\end{eqnarray}%
which have the parameters $\lambda _{i}$, $i=1,2$. In terms of these
symmetries, one can rescale the trial solution to meet the UV boundary
condition $f=1$ and $A=r/L$.

\begin{figure}[th]
\centerline{
\includegraphics[width=.8\textwidth]{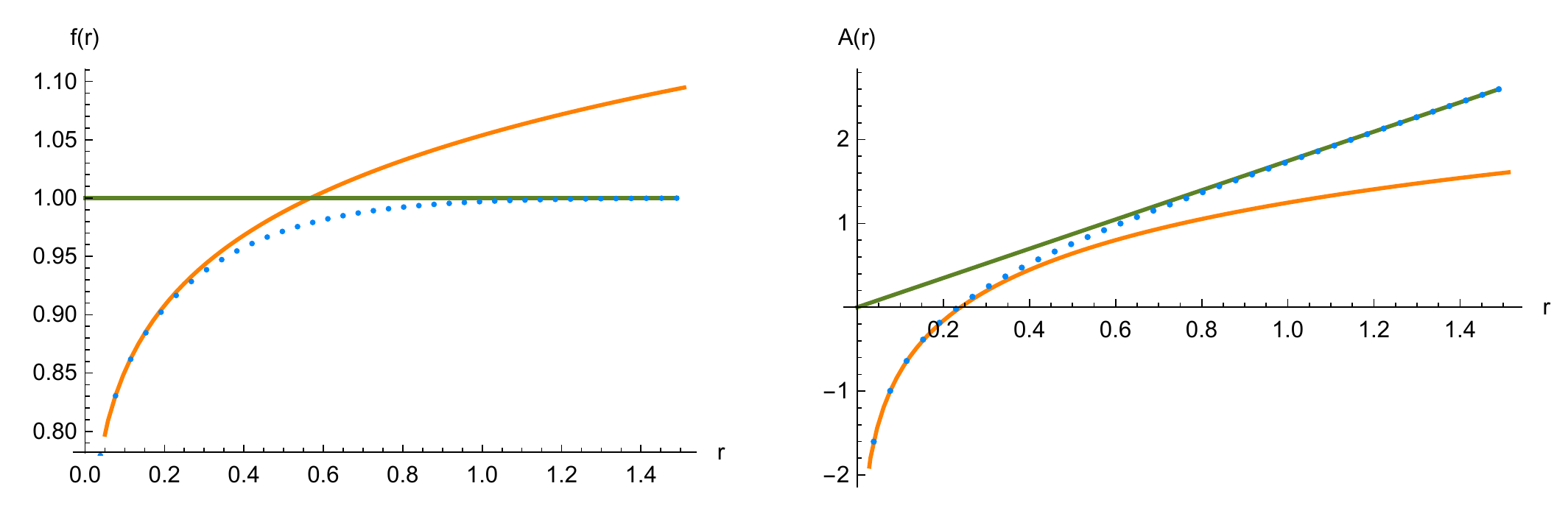}}
\caption{The metric functions $f(r)$ (left) and $A(r)$ (right) in the FQH
model. They are dotted lines and coloured in blue. The orange line and the
green line denote the IR and UV asymptotics, respectively.}
\label{FIGFQH2}
\end{figure}

As an example, we will construct the trial solution by setting $\tau _{\ast
}=e^{\frac{\pi i}{3}}$ and selecting the parameters%
\begin{equation}
\gamma =-0.85,\;s=1.2,\;a_{0}=0.2,\;p=1,\;q=1,\;r_{0}=0.  \label{para}
\end{equation}%
The final metric functions are plotted in Figure \ref{FIGFQH2}.

\section{Gauss-Bonnet formula on three-dimensional manifolds}

We further explore the HSC in the GB gravity. Since we suspect that Eq. (\ref%
{CHD}) might need to be improved, a natural question is what the expected
expression looks like.

Let's write down the HEE in the GB gravity \cite%
{Fursaev0606,Boer1101,Hung1101}%
\begin{equation}
\mathcal{S}=\frac{1}{4G_{N}}\left[ \int_{\Sigma }d^{2}x\sqrt{h}+\alpha
\left( \int_{\Sigma }d^{2}x\sqrt{h}\;^{(2)}R+2\int_{\partial \Sigma }dx\sqrt{%
h_{\partial }}K\right) \right] ,  \label{SGB}
\end{equation}%
where $h$ is the determinant of the metric on the RT surface $\Sigma $, $%
^{(2)}R$ is the intrinsic Ricci scalar defined on $\Sigma $, $h_{\partial }$
is the determinant of induced metric on the boundary $\partial \Sigma $, and
$K$ is the extrinsic curvature of the boundary. The terms in the parentheses
are proportional to the Euler characteristic $\chi $ of the RT surface due
to the GB theorem.

We turn to observe the proposals for the HSC in the GB gravity, see Eq. (\ref%
{CHD0}) and Eq. (\ref{CHD}). They are mainly motivated by the expression of
the HEE (or Wald entropy) in the higher-derivative theory. However,
comparing them with Eq. (\ref{SGB}), one can find an obvious difference:
there are no topological features in Eq. (\ref{CHD0}) and Eq. (\ref{CHD}).
In view of this, a natural proposal for the HSC in the GB gravity would be
to involve a topological invariant like the integral of the Euler density,
but it should be defined on the three-dimensional space inside the RT
surface. It is well-known that the even dimensionality is essential for the
GB theorem. If the manifold is odd-dimensional, the Euler characteristic is
zero. Interestingly, we notice that there is a formula in the
odd-dimensional manifold which is actually related to the GB theorem \cite%
{Tanno1972}:%
\begin{equation}
\frac{1}{l(\xi )}\frac{1}{2\pi }\left[ \int_{M}\mathrm{K}dM+3\mathrm{Vol}(M)%
\right] =2\beta _{0}\left( M\right) -\beta _{1}\left( M\right) ,
\end{equation}%
where $M$ is a three-dimensional Sasakian manifold, $\xi $ is the regular
unit Killing vector, $\mathrm{K}$ is the sectional curvature of the surface
orthogonal to $\xi $, $l(\xi )$ is the length of its trajectory, and $\beta
_{i}\left( M\right) $ denotes the i-th Betti number of $M$. In particular,
the length scale seems to appear in a suitable place. However, the space
inside the RT surface is not the Sasakian manifold in general. In the
future, it would be interesting to investigate whether the restriction on
the manifold could be\ sufficiently relaxed\footnote{%
See relevant discussion in \cite{Reventos1979}.} and $l(\xi )$ could be
related to the size of entangling surface. We emphasize that this direction
is presently quite speculative.

\end{document}